\documentclass[fleqn,usenatbib]{mnras}

\usepackage{newtxtext,newtxmath}

\usepackage[T1]{fontenc}
\usepackage{ae,aecompl}

\usepackage{graphicx}	
\usepackage{amsmath}	
\usepackage{amssymb}	
\usepackage{hyperref}
\hypersetup{draft}
\usepackage{subfig}
\hypersetup{final}            

\title[\textit{CXO} and \textit{HST} observations of dark GRB hosts]{\textit{Chandra} and \textit{Hubble Space Telescope} observations of dark gamma-ray bursts and their host galaxies}

\author[A. A. Chrimes et al.]{A. A. Chrimes,$^{1}$\thanks{E-mail: A.Chrimes@warwick.ac.uk}
A. J. Levan,$^{1}$
E. R. Stanway,$^{1}$
J. D. Lyman,$^{1}$
A. S. Fruchter,$^{2}$ \newauthor
P. Jakobsson,$^{3}$
P. O'Brien,$^{4}$
D. A. Perley,$^{5}$
N. R. Tanvir,$^{4}$
P. J. Wheatley$^{1}$\newauthor and
K. Wiersema$^{1}$
\\
$^{1}$Department of Physics,  University of Warwick, Gibbet Hill Road, Coventry, CV4 7AL, UK\\
$^{2}$Space Telescope Science Institute, 3700 San Martin Drive, Baltimore, MD21218, USA\\
$^{3}$Centre for Astrophysics and Cosmology, Science Institute, University of Iceland, Dunhagi 5, 107 Reykjavik, Iceland\\
$^{4}$Department of Physics and Astronomy, University of Leicester, University Road, Leicester, LE1 7RH, UK\\
$^{5}$Astrophysics Research Institute, Liverpool John Moores University, 146 Brownlow Hill, Liverpool L3 5RF, UK\\
}

\date{Accepted XXX. Received YYY; in original form ZZZ}

\pubyear{2019}

\begin{document}
\label{firstpage}
\pagerange{\pageref{firstpage}--\pageref{lastpage}}
\maketitle

\begin{abstract}
We present a study of 21 dark gamma-ray burst (GRB) host galaxies, predominantly using X-ray afterglows obtained with the {\it Chandra X-Ray Observatory} ({\it CXO}) to precisely locate the burst in deep {\it Hubble Space Telescope} ({\it HST}) imaging of the burst region. 
The host galaxies are well-detected in F160W in all but one case and in F606W imaging in $\sim 60$ per cent of cases. We measure magnitudes and perform a morphological analysis of each galaxy. The asymmetry, concentration and ellipticity of the dark burst hosts are compared against the host galaxies of optically bright GRBs. In agreement with other studies, we find that dark GRB hosts are redder and more luminous than the bulk of the GRB host population. The  distribution of projected spatial offsets for dark GRBs from their host galaxy centroids is comparable to that of optically-bright bursts. The dark GRB hosts are physically larger, more massive and redder, but are morphologically similar to the hosts of bright GRBs in terms of concentration and asymmetry.
Our analysis constrains the fraction of high redshift ($z>5$) GRBs in the sample to ${\sim}$14 per cent, implying an upper limit for the whole long-GRB population of $\le$4.4 per cent.  If  dust is the primary cause of afterglow darkening amongst dark GRBs, the measured  extinction may require a clumpy dust component in order to explain the observed offset and ellipticity distributions. 
\end{abstract}

\begin{keywords}
gamma-ray burst: general, galaxies: photometry, galaxies: ISM, galaxies: structure, galaxies: high-redshift
\end{keywords}



\section{Introduction}
Long-duration Gamma-ray bursts (GRBs)\footnote{Hereafter we use GRB to refer to long GRBs, and there is no discussion of the short duration bursts.} are the most luminous events in the Universe \citep[e.g.][]{2008Natur.455..183R}, arising from the violent explosions of massive stars \citep[e.g.][]{2003Natur.423..847H}. Newly formed compact objects can launch strongly beamed relativistic jets, producing the prompt gamma-ray emission. As the jet expands and the ejecta cools, it interacts with the circumstellar medium, producing external shocks which manifest as an afterglow. The wavelength of peak afterglow emission increases over time, with the spectral shape well described in most cases by a synchrotron-like broken power law \citep{1993ApJ...413..281B,1994ApJ...432..181M,1998ApJ...497L..17S}. While most GRBs display such an afterglow in the optical bands if deep and early follow-up imaging is performed, a significant minority do not. The first example where this was found to be the case was GRB\,970828 \citep{1998ApJ...502L.123G}, which showed no afterglow down to an $r$-band limit of AB mag ${\sim}$ 23 within 12 hours post-burst. Such events have subsequently become known as dark bursts \citep{2001A&A...369..373F}. A commonly used formal definition for dark GRBs is an X-ray to optical spectral slope ${\beta}_{\mathrm{OX}}$ of less than ${\sim}$ 0.5 \citep{2004ApJ...617L..21J}, effectively the limit allowed by standard synchrotron afterglow theory. Alternatively, the X-ray spectral slope can be extrapolated \citep{2005ApJ...624..868R} according to a power of the form $F_{\nu}\,{\alpha}\,{\nu}^{-{\beta_{X}}}$, and darkness defined as when ${\beta}_{\mathrm{OX}} < {\beta}_{\mathrm{X}} - 0.5$ \citep{2009ApJ...699.1087V}. It should be noted however that GRB emission can deviate from this simple synchrotron model, with plateaus, flares and variable decay rates often being seen \citep{2009MNRAS.397.1177E,0004-637X-866-2-162}. Estimates for the fraction of GRBs which are dark vary, but are typically around 25-40 per cent \citep[e.g.][]{2009ApJS..185..526F,2011A&A...526A..30G,2016ApJ...817....7P}.

There are three possible explanations for darkness in GRBs \citep{2011A&A...526A..30G}. Firstly, the burst may be intrinsically suppressed at optical wavelengths. Although this is disfavoured due to the difficulty in explaining such a spectral shape, it may be plausible in particularly low density environments, or if the spectral energy distribution is measured during a non-standard phase such as a flare or plateau. Second, the burst may be at high redshift (we define this as $z>5$), where observations in the optical correspond to rest frame wavelengths blue-wards of the Lyman break (noting also that the L${\alpha}$ forest may be dense enough to produce a comparable effect at redshift $4<z<5$). GRBs at $z>5$ are known to be rare in the spectroscopically confirmed sample \citep[e.g.][]{2006Natur.440..184K,2007ApJ...669....1R,Salvaterra,2009ApJ...693.1610G,2009Natur.461.1254T,2011ApJ...736....7C,Tanvir_hi-z}. Finally, the host galaxy (or Milky Way sight-line) might be dusty, so that the optical afterglow is reddened and attenuated. The last of these is favoured as the most frequent scenario, not least because the host galaxies of dark bursts are often detected at optical wavelengths, ruling out a high redshift origin. The inferred rate of GRBs at high redshift is therefore low \citep[current estimates put ${\sim}$10-20 per cent of dark GRBs at $z$ > 5,][]{2011A&A...526A..30G,2012ApJ...752...62J,2016ApJ...817....7P}.

GRBs are known to arise from the collapse of rapidly rotating, massive stars from their association with broad line type Ic supernovae (SNe) \citep{1993ApJ...405..273W,1999A&AS..138..499W,2012grb..book..169H,2017AdAst2017E...5C}. Beyond this, however, their production mechanisms and progenitors are not well understood \citep{2016SSRv..202...33L}. The study of GRB host galaxies has provided additional insight into the environments capable of producing GRBs \citep{2002MNRAS.329..465R,2002MNRAS.334..983T}, and by extension the nature of the progenitor systems. The GRB host population is overwhelmingly star forming and the burst locations trace this star formation, as measured through both projected, host normalised offsets and the fractional light $F_{\mathrm{light}}$ statistic \citep{2002AJ....123.1111B,2006Natur.441..463F,2010MNRAS.405...57S,2016ApJ...817..144B,2017MNRAS.467.1795L,2018A&A...617A.105J}. 

GRBs do not appear to be entirely unbiased tracers of star formation, however. Early studies of GRB hosts reported a strong bias against massive galaxies, implying some level of metallicity aversion in GRBs \citep[e.g.][]{2006Natur.441..463F} due to the mass metallicity relation \citep{2004ApJ...613..898T}. However, the first studies of this kind tended to use optical afterglows for host localisation, and therefore systematically omitted the hosts of dark GRBs from their samples. Subsequently, efforts have been made to account for this effect by specifically including dark hosts \citep{2009ApJ...693.1484C,2011A&A...534A.108K,2012ApJ...756..187H,2013ApJ...778..128P,2016ApJ...817....7P}, made possible by NIR afterglow imaging, or X-ray facilities such as the {\it Neil Gehrels Swift Observatory} \citep{Swift} and its on-board X-Ray Telescope \citep[XRT; ][]{XRT}. XRT provides ${\gtrsim}$ 1 arcsec localisation accuracy, sufficient in some cases to identify a probable host candidate. Because most dark GRBs are hosted by galaxies which are more massive, dustier and more chemically enriched than the wider population, their inclusion should weaken any bias relative to the underlying star formation distribution. Optically unbiased GRB host studies have shown this to be true, but despite the addition of more massive GRB hosts, some form of metallicity bias still appears to exist in the population \citep{2013ApJ...778..128P,2015A&A...581A.125K,2016ApJ...817....7P,2016ApJ...817....8P}. However, the precise value of this cutoff remains uncertain. There are a handful of cases with ${\sim}$ solar metallicity which suggest a hard cut-off is unlikely \citep{2015arXiv151100667G,2017ApJ...834..170G}. If GRBs can genuinely be 
created at solar metallicity, it is challenging for single star progenitor models which predict too much mass and angular momentum loss through winds at these metallicities \citep{2001A&A...369..574V,2005A&A...443..581H}. Solutions have been offered in the form of chemically homogeneous evolution, or binary pathways \citep[e.g.][ and references therein]{2012A&A...542A.113Y,2015A&A...581A..15S,2016A&A...585A.120S,2017PASA...34...58E}. The exact nature of the host galaxy bias is still debated, with implications for both the progenitors and the usefulness of GRBs as tracers of star formation across cosmic time. 

In this paper, we present a study of 21 dark GRBs and their host galaxies, observed with the {\it Chandra X-ray Observatory} ({\it CXO}) and {\it Hubble Space Telescope} ({\it HST}). The sub arcsecond astrometric accuracy of {\it CXO} X-ray imaging, combined with deep {\it HST} optical and NIR imaging, allows us to precisely locate the bursts and identify faint hosts down to AB mag ${\sim}$ 27. 

As well as increasing the statistical certainty that dark GRBs favour luminous, dusty hosts, the spatial resolution of {\it HST} allows us to examine the projected morphology of the GRB hosts. Crucially, these data also allow us to put constraints on the fraction of dark bursts arising from high redshift ($z>5$). The paper is structured as follows. In section \ref{sec:obs}, we detail the observations and data reduction. Section \ref{sec:method} outlines the methodology, and in section \ref{sec:results} we present our results. This is followed by the discussion and conclusions in sections \ref{sec:discuss} and \ref{sec:conc}. Throughout, magnitudes are quoted in the AB system \citep{1983ApJ...266..713O}. A flat $\Lambda$CDM cosmology with $h$ = 0.7, ${\Omega}_\mathrm{M}$ = 0.3 and ${\Omega}_{\Lambda}$ = 0.7 is used.


\section{Observations and Data Reduction}\label{sec:obs}

\subsection{Target Catalogue}\label{sec:target}

A total of 21 dark GRB positions were imaged with {\it HST} (PI: Levan)\footnote{Programmes 11343, 11840, 12378, 12764, 13117 and 13949}. The criteria for inclusion was an X-ray to optical spectral slope, ${\beta}_{\mathrm{OX}}$, of less than 0.5 (within 12 hours post-burst), and a Galactic foreground extinction of $A_V<0.5$ \citep[determined from the dust maps of ][]{1998ApJ...500..525S}. For {\em CXO} observations it was necessary that no more precise position (e.g. optical/radio) was available at the time of the {\em CXO} trigger. No further selection criteria were applied, although not all candidates in a given cycle could be followed up due to limits on the available observing time. For each burst, a ${\beta}_{\mathrm{OX}}$ limit is provided in table \ref{tab:Box}. Where an analysis has not already been performed in the literature, these are determined from reported optical limits, and the extrapolated X-ray flux at the time of these observations assuming a simple power law\footnote{\url{http://www.swift.ac.uk/xrt_curves}}. 

\begin{table}
\centering 
\caption{Approximate ${\beta}_{\mathrm{OX}}$ limits, calculated by extrapolation of the X-ray lightcurves out to the time of deep optical observations. We correct for Galactic foreground extinction with the updated dust maps of \citet{2011ApJ...737..103S}. Otherwise, where a detailed analysis of the afterglow has been carried out in the literature, that value is reported here.} 
\begin{tabular}{l c p{5cm}}
\hline 
GRB & ${\beta}_{\mathrm{OX}}$ & Reference \\ 
\hline 
051022$^{\dagger}$	&	$<$-0.1	&	\citet{2007ApJ...669.1098R}	\\
080207$^{\dagger}$	&	$<$0.3	&	\citet{2012MNRAS.421...25S}	\\
090113$^{\dagger}$	&	$<$0.3	&	\citet{2012ApJ...758...46K}	\\
090404$^{\dagger}$	&	$<$0.2	&	\citet{2013ApJ...778..128P}	\\
090407$^{\dagger}$	&	$<$0.4	&	\citet{2012ApJ...758...46K}	\\
090417B$^{\dagger}$	&	$<$-1.9	&	\citet{2010ApJ...717..223H}	\\
100205A	&	$<$0.3	&	\citet{2010GCN.10362....1M}	\\
100413A	&	$<$0.2	&	\citet{2010GCN.10592....1F}	\\
100615A &	$<$-0.6	&	\citet{2010GCN.10844....1N}	\\
110312A	&	$<$0.2	&	\citet{2011GCN.11785....1N}	\\
110709B	&	$<$-0.1	&	\citet{2011GCN.12155....1F}	\\
110915A	&	$<$0.2	&	\citet{2011GCN.12343....1M}	\\
111215A$^{\dagger}$	&	$<$0.2	&	\citet{2015MNRAS.446.4116V}	\\
120320A &	$<$0.5	&	\citet{2012GCN.13083....1C}	\\
130131A &	$<$0.4	&	\citet{2013GCN.14180....1S}	\\
130502B	&	$<$0.3	&	\citet{2013GCN.14552....1M}	\\
130803A	&	$<$0.5	&	\citet{2013GCN.15067....1L}	\\
131229A	&	$<$-0.3	&	\citet{2013GCN.15636....1G}	\\
140331A &	$<$0.2	&	\citet{2014GCN.16076....1B}	\\
141031A	&	$<$0.1	&	\citet{2014GCN.17004....1T}	\\
150616A	&	$<$-0.4	&	\citet{2015GCN.17947....1M}	\\
\hline 
\end{tabular}
\newline
${\dagger}$ - These ${\beta}_{\mathrm{OX}}$ limits are obtained from the literature.
\label{tab:Box}
\end{table}

\begin{figure}
\centering
\includegraphics[width=0.9\columnwidth]{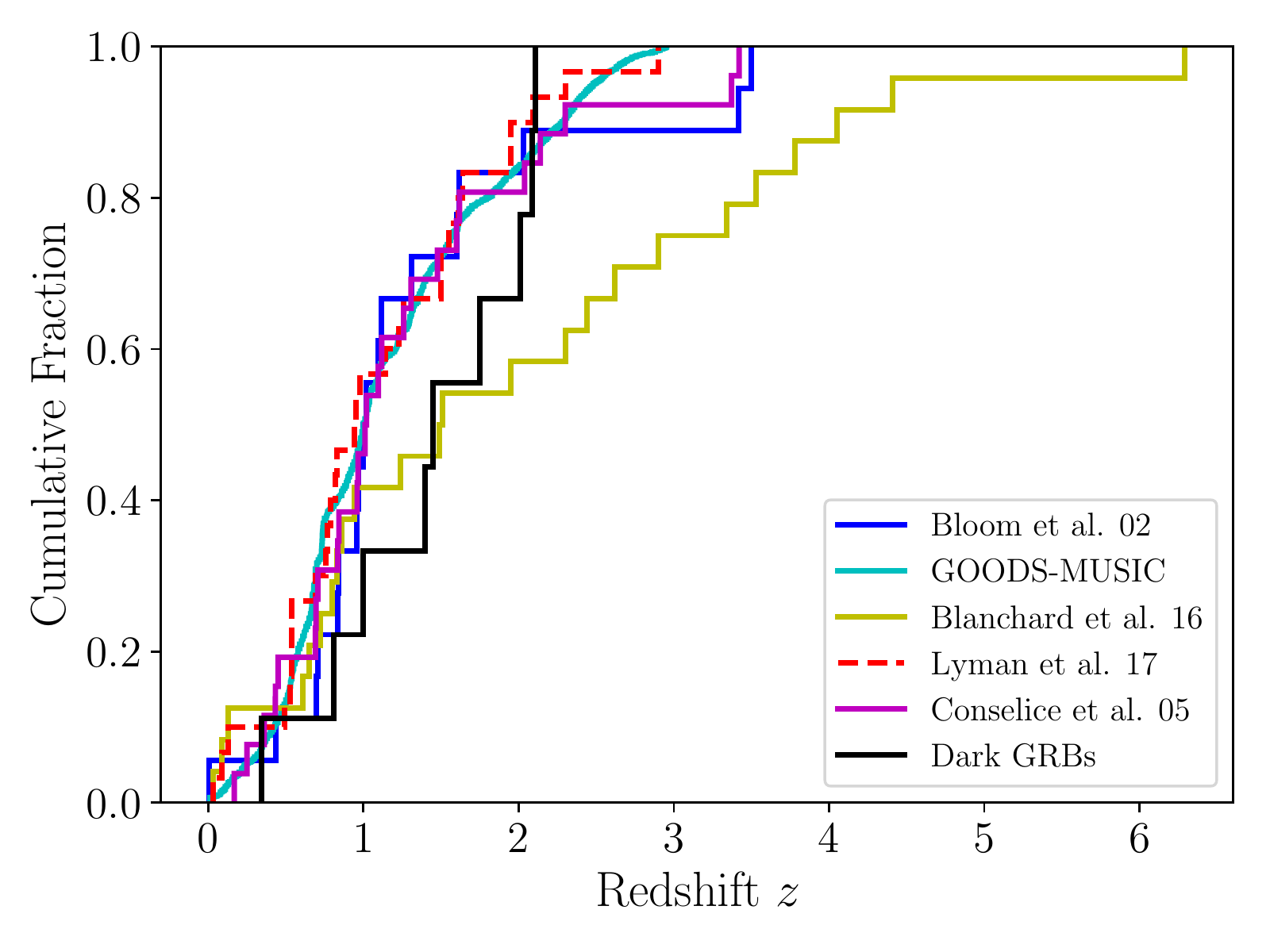}
\caption{A comparison between the redshift distribution of the dark GRB sample, other similar studies and the GOODS-MUSIC galaxy survey. The dark hosts with redshifts have a distribution comparable to the other samples, however many of them do not have this information and may be biased towards higher $z$.}
\label{fig:redshifts}
\end{figure}

Because the optical afterglows of dark GRBs are by definition faint or undetected, they seldom yield absorption-line redshifts. Redshifts for dark bursts must therefore come from observations of a likely host candidate. Only 9 of the sample have redshifts (either photometric or spectroscopic) from the literature, these are listed in table \ref{tab:needsz}. In figure \ref{fig:redshifts}, we compare the {\it known} redshifts in this sample to the redshift distributions from \citet{2002AJ....123.1111B}, \citet[][ with which there is sample overlap]{2016ApJ...817..144B} and \citet{2017MNRAS.467.1795L}, who all provide burst-host galaxy spatial offsets for mixed (dark and bright) or exclusively optically-bright GRB samples. We also show the redshift distribution of \citet{2005ApJ...633...29C}, whose concentrations and asymmetries we compare to later, and the GOODS-MUSIC galaxy survey \citep{2006A&A...449..951G,2009yCat..35040751S}. In all cases, the distributions are similar, and assuming that dark GRBs without redshifts are not significantly biased towards high $z$, these therefore represent fair comparison samples for the parameters of interest.

\subsection{\textit{Hubble Space Telescope}}
Each burst location was imaged with {\it HST} in two bands, F160W ($\lambda_\mathrm{eff} \sim 15400$\AA, IR) and F606W ($\lambda_\mathrm{eff} \sim 6060$\AA, UVIS). An exception is GRB\,080207 which has F110W imaging instead of F160W for the IR \citep{2012MNRAS.421...25S}. For all IR observations, and most UVIS, the Wide Field Camera 3 (WFC3) was used. For 4 UVIS observations, the Advanced Camera for Surveys (ACS) was employed, and the Wide Field Planetary Camera 2 (WFPC2) was used once (for GRB\,080207). The details of these observations are given in table \ref{tab:HSTobservations}. 

The exposures for the {\it HST} targets were dithered, at least twice or up to 4 times depending on the exposure times. The charge transfer efficiency (CTE) corrected images were reduced with standard {\sc Astrodrizzle} procedures, available with the python package {\sc drizzlepac}\footnote{\url{http://drizzlepac.stsci.edu}}. The {\sc pixfrac} was chosen to be 0.8 in every case, while the final scale is 0.065 arcsec pixel$^{-1}$ for IR images and 0.02 arcsec pixel$^{-1}$ for UVIS. Exceptions are the ACS images where we use a 0.03 arcsec pixel$^{-1}$ final scale, the sole WFPC2 example where it is 0.07 arcsec pixel$^{-1}$, and the three IR images in programme 13949 where only two dithers were available and the final scale is 0.085 arcsec pixel$^{-1}$. A subset of these data were previously published in \citet{2016ApJ...817..144B}, and we obtain similar results in these cases.


\begin{table}
\centering 
\caption{Details of the {\it HST} observations.} 
\begin{tabular}{l c l c c c c}
\hline 
GRB & Prog. & Date & Inst. & Filter & Exp. [s] \\ 
\hline 
051022	&	11343	&	2009 Oct 12	&	WFC3	&	F160W	&	2397		\\
051022	&	11343	&	2009 Aug 21	&	ACS	&	F606W	&	2080		\\
080207	&	11343	&	2009 Dec 09	&	WFC3	&	F110W	&	1600		\\
080207	&	11343	&	2008 Mar 18	&	WFPC2	&	F606W	&	1600		\\
090113	&	11840	&	2009 Oct 17	&	WFC3	&	F160W	&	2612		\\
090113	&	11840	&	2009 Oct 15	&	ACS	&	F606W	&	2208		\\
090404	&	11840	&	2010 Jan 09	&	WFC3	&	F160W	&	2612		\\
090404	&	11840	&	2010 Sep 02	&	ACS	&	F606W	&	2208		\\
090407	&	11840	&	2010 Sep 15	&	WFC3	&	F606W	&	740		\\
090407	&	11840	&	2010 Sep 15	&	WFC3	&	F160W	&	1209		\\
090417B	&	11840	&	2009 Oct 17	&	WFC3	&	F160W	&	2612		\\
090417B	&	11840	&	2011 Jan 22	&	ACS	&	F606W	&	1656		\\
100205A	&	11840	&	2010 Dec 06	&	WFC3	&	F606W	&	1140		\\
100205A	&	11840	&	2010 Dec 06	&	WFC3	&	F160W	&	1209		\\
100413A	&	11840	&	2010 Aug 31	&	WFC3	&	F606W	&	752		\\
100413A	&	11840	&	2010 Aug 31	&	WFC3	&	F160W	&	1209		\\
100615A	&	11840	&	2010 Dec 16	&	WFC3	&	F606W	&	1128		\\
100615A	&	11840	&	2010 Dec 16	&	WFC3	&	F160W	&	1209		\\
110312A	&	12378	&	2011 Nov 17	&	WFC3	&	F606W	&	1110		\\
110312A	&	12378	&	2011 Nov 18	&	WFC3	&	F160W	&	1209		\\
110709B	&	12378	&	2011 Nov 12	&	WFC3	&	F160W	&	2612		\\
110709B	&	12378	&	2011 Nov 08	&	WFC3	&	F606W	&	2480		\\
110915A	&	12764	&	2011 Nov 03	&	WFC3	&	F160W	&	2612		\\
110915A	&	12764	&	2011 Oct 31	&	WFC3	&	F606W	&	2508		\\
111215A	&	12764	&	2013 May 13	&	WFC3	&	F160W	&	1209		\\
111215A	&	12764	&	2013 May 13	&	WFC3	&	F606W	&	1110		\\
120320A	&	12764	&	2013 Feb 20	&	WFC3	&	F606W	&	1110		\\
120320A	&	12764	&	2013 Feb 20	&	WFC3	&	F160W	&	1209		\\
130131A	&	13117	&	2014 Oct 09	&	WFC3	&	F160W	&	1059		\\
130131A	&	13117	&	2014 Oct 09	&	WFC3	&	F606W	&	1101		\\
130502B	&	13117	&	2013 Dec 30	&	WFC3	&	F160W	&	2412		\\
130502B	&	13117	&	2013 Dec 30	&	WFC3	&	F606W	&	2400		\\
130803A	&	13117	&	2014 May 28	&	WFC3	&	F160W	&	1209		\\
130803A	&	13117	&	2014 May 28	&	WFC3	&	F606W	&	1125		\\
131229A	&	13117	&	2014 Aug 14	&	WFC3	&	F160W	&	1209		\\
131229A	&	13117	&	2014 Aug 14	&	WFC3	&	F606W	&	1125		\\
140331A	&	13949	&	2016 Mar 28	&	WFC3	&	F160W	&	1209		\\
140331A	&	13949	&	2016 Mar 28	&	WFC3	&	F606W	&	1137		\\
141031A	&	13949	&	2014 Nov 29	&	WFC3	&	F160W	&	1209		\\
141031A	&	13949	&	2014 Nov 29	&	WFC3	&	F606W	&	1395		\\
150616A	&	13949	&	2016 Feb 29	&	WFC3	&	F160W	&	1209		\\
150616A	&	13949	&	2016 Feb 29	&	WFC3	&	F606W	&	1329		\\
\hline 
\end{tabular}
\label{tab:HSTobservations}
\end{table}

\subsection{\textit{Chandra}}
Out of 21 burst locations observed with {\it HST}, 18 have been observed with {\it CXO} and its Advanced CCD Imaging Spectrometer (ACIS) instrument (all PI: Levan, with the exception of GRB\,051022, PI: Kouveliotou). Standard {\sc ciao} (v4.9, with {\sc caldb} v4.7.6) procedures were used to reduce the data, including reprocessing, PSF map creation and energy filtering of the event files to the range 0.3-8\,keV. {\sc wavdetect} is then used to identify sources in the field. A list of the observations used is provided in table \ref{tab:CXOobservations}.

\begin{table*}
\centering 
\caption{Details of the {\it CXO} observations. All imaging was performed with ACIS-S. The modified Julian Date (MJD) of the observation mid-point is provided, as is the mean count rate in the 0.3-8\,keV energy range. The J2000 R.A. and Dec of the afterglow, in the {\it CXO} world coordinate system, is also listed.} 
\begin{tabular}{l c l l c c l l}
\hline 
GRB & Obsv. ID & Date & MJD & Exp. [ks] & Count Rate [s$^{-1}$] & R.A. & Dec. \\ [0.25ex] 
\hline 
051022	&	5536	&	2005 Oct 05	&	53668.88	&	18.72	&	(2.83${\pm}$0.03)${\times}$10$^{-2}$	&	23:56:04.09	&	+19:36:23.90	\\
080207	&	9474	&	2008 Feb 15	&	54511.97	&	14.83	&	(6.00${\pm}$0.93)${\times}$10$^{-4}$	&	13:50:02.97	&	+07:30:07.60	\\
090113	&	10490	&	2009 Jan 19	&	54850.52	&	14.85	&	(2.09${\pm}$0.17)${\times}$10$^{-3}$ &	02:08:13.77	&	+33:25:43.30	\\
090404	&	10491	&	2009 Apr 13	&	54934.35	&	14.85	&	(6.53${\pm}$0.22)${\times}$10$^{-3}$ &	15:56:57.50	&	+35:30:57.50	\\
090407	&	10492	&	2009 Apr 18	&	54939.79	&	14.96	&	(2.66${\pm}$0.17)${\times}$10$^{-3}$ &	04:35:55.06	&	\textminus12:40:45.10	\\
090417B	&	10493	&	2009 May 11	&	54962.67	&	13.70	&	(1.31${\pm}$0.15)${\times}$10$^{-3}$ &	13:58:46.63	&	+47:01:04.40	\\
100413A	&	11772	&	2010 Apr 19	&	55305.05	&	14.94	&	(3.51${\pm}$0.18)${\times}$10$^{-3}$ &	17:44:53.13	&	+15:50:03.70	\\
100615A	&	12229	&	2010 Jun 21	&	55368.15	&	14.84	&	(1.48${\pm}$0.03)${\times}$10$^{-2}$ &	11:48:49.34	&	\textminus19:28:52.00	\\
110312A	&	12919	&	2011 Mar 22	&	55642.95	&	14.86	&	(7.63${\pm}$0.22)${\times}$10$^{-3}$ &	10:29:55.49	&	\textminus05:15:44.70	\\
110709B	&	12921	&	2011 Jul 23	&	55765.59	&	14.86	&	(4.18${\pm}$0.19)${\times}$10$^{-3}$ &	10:58:37.11	&	\textminus23:27:16.90	\\
110915A	&	14051	&	2011 Sep 26	&	55830.66	&	14.86	&	(1.25${\pm}$0.15)${\times}$10$^{-3}$ &	20:43:17.93	&	\textminus00:43:23.90	\\
111215A	&	14052	&	2011 Dec 28	&	55923.20	&	14.77	&	(9.22${\pm}$0.23)${\times}$10$^{-3}$ &	23:18:13.30	&	+32:29:39.40	\\
120320A	&	14053	&	2012 Mar 26	&	56012.55	&	15.07	&	(4.62${\pm}$0.13)${\times}$10$^{-3}$ &	14:10:04.27	&	+08:41:47.60	\\
130502B	&	15194	&	2013 May 13	&	56425.07	&	14.69	&	(2.70${\pm}$0.18)${\times}$10$^{-3}$ &	04:27:03.07	&	+71:03:36.50	\\
131229A	&	15195	&	2014 Jan 06	&	56663.12	&	15.05	&	(1.06${\pm}$0.15)${\times}$10$^{-3}$ &	05:40:55.62	&	\textminus04:23:46.50	\\
140331A	&	16161	&	2014 Apr 08	&	56755.15	&	14.86	&	(5.52${\pm}$1.25)${\times}$10$^{-4}$ &	08:59:27.51	&	+02:43:02.80	\\
141031A	&	16162	&	2014 Nov 06	&	56967.30	&	10.19	&	(4.58${\pm}$0.26)${\times}$10$^{-3}$ &	08:34:26.09	&	\textminus59:10:05.80	\\
150616A	&	17235	&	2016 Jun 24	&	57197.30	&	14.76	&	(8.63${\pm}$0.24)${\times}$10$^{-3}$ &	20:58:52.00	&	\textminus53:23:38.00	\\
\hline 
\end{tabular}
\label{tab:CXOobservations}
\end{table*}

\section{Methodology}\label{sec:method}
\subsection{Astrometric Alignment}
The default pipeline processing can result in astrometric offsets between {\it Chandra} and {\it HST} of order a few arcsec. Because we expect the burst-host offsets to be much smaller than this \citep[e.g,][]{2016ApJ...817..144B,2017MNRAS.467.1795L}, a refined astronometric solution was required to precisely locate the burst with respect to the host. This involved identifying sources in common between images, and computing the best transformation that maps one set of coordinates onto the other, a process referred to as astrometric tying. In almost all cases, there were insufficient sources in common between {\it HST} and {\it CXO} to perform a direct tie. Instead, an intermediate was used, which was in most cases a Pan-STARRS\footnote{The Panoramic Survey Telescope and Rapid Response System, see \url{http://panstarrs.stsci.edu}} cutout \citep{2016arXiv161205560C}. Again, due to the low number of sources detected by {\it CXO} and the faintness of their associated optical counterparts, there were only a handful of {\it CXO}-Pan-STARRS matches in each case. To tie these images, we performed a similarity transform on the {\it CXO} coordinates, placing them in the intermediate frame. This transform conserves the relative distances between points, and involves an $x-y$ shift, scaling and rotation. The scaling between images was known and fixed, and if necessary to avoid over-fitting, the rotation obtained from the image headers was assumed to be correct. In this scenario, the root-mean-square uncertainties, whilst incorporating offsets due to any rotation errors present, were derived as if only $x-y$ shifts were contributing. In this way, rotational uncertainties are still accounted for.


The next step, tying Pan-STARRS or an alternative intermediate to ${\it HST}$, provided many more tie objects allowing for a more sophisticated procedure. Tying was performed with the {\sc IRAF} tasks {\sc geomap} and {\sc geoxytran}, fitting for rotation, scaling in $x$ and $y$, $x$-$y$ shifts, and second or third order polynomial distortions.

The total tie uncertainty was estimated as the quadrature sum of the X-ray to intermediate, and intermediate to {\it HST} root-mean-square uncertainties. In turn, this was added in quadrature to the afterglow positional uncertainty, which is estimated as FWHM/(2.35\,SNR) \citep[where FWHM is the full-width at half maximum, and SNR is the signal-to-noise ratio, e.g.][]{2006obas.book.....B}. When measuring burst offsets, the uncertainty on the host centre was also considered, however this was usually at the sub-pixel level. Deviations from these standard procedures are as follows:
\newline
\textit{GRB100205A}: No {\it CXO} data was available for this burst. Instead, the {\it HST} images were tied directly to a $K$-band Gemini-North image of the afterglow \citep{2010GCN.10366....1T,2010GCN.10374....1C}. There was therefore no need for an intermediate image in this case.
\newline
\textit{GRB130131A}: No {\it CXO} imaging is available. The {\it HST} images are tied directly to a $K$-band UKIRT image of the afterglow \citep{2013GCN.14175....1T}. Again, no intermediate image was required.
\newline
\textit{GRB130803A}: No {\it CXO}, optical or NIR afterglow was available, so we used the less-precise enhanced XRT position\footnote{\url{www.swift.ac.uk/xrt_positions}}. 
\newline
\textit{GRB141031A}: This source was too far south for Pan-STARRS coverage, and there were insufficient mutually detected objects between the {\it CXO} image and other available intermediates (e.g. 2MASS, SDSS) to perform a tie. Instead, we directly placed the {\it CXO} position onto the {\it HST} frame, after it had been astrometrically refined through cross-matching to the Hubble Source Catalogue \citep[v3,][]{2016AJ....151..134W}. The uncertainty on the burst position in the {\it HST} images was therefore given by the absolute astrometric accuracy of both {\it CXO} and the refined {\it HST} image, as well as the positional uncertainty of the {\it CXO} afterglow. 

\textit{GRB150616A}: This burst was too far south for Pan-STARRS coverage. Instead, we used a VLT/FORS2 image as the intermediate, from programme 095.B-0811(C) (PI: Levan). 

\begin{figure*}
\centering
\includegraphics[height=0.9\textheight]{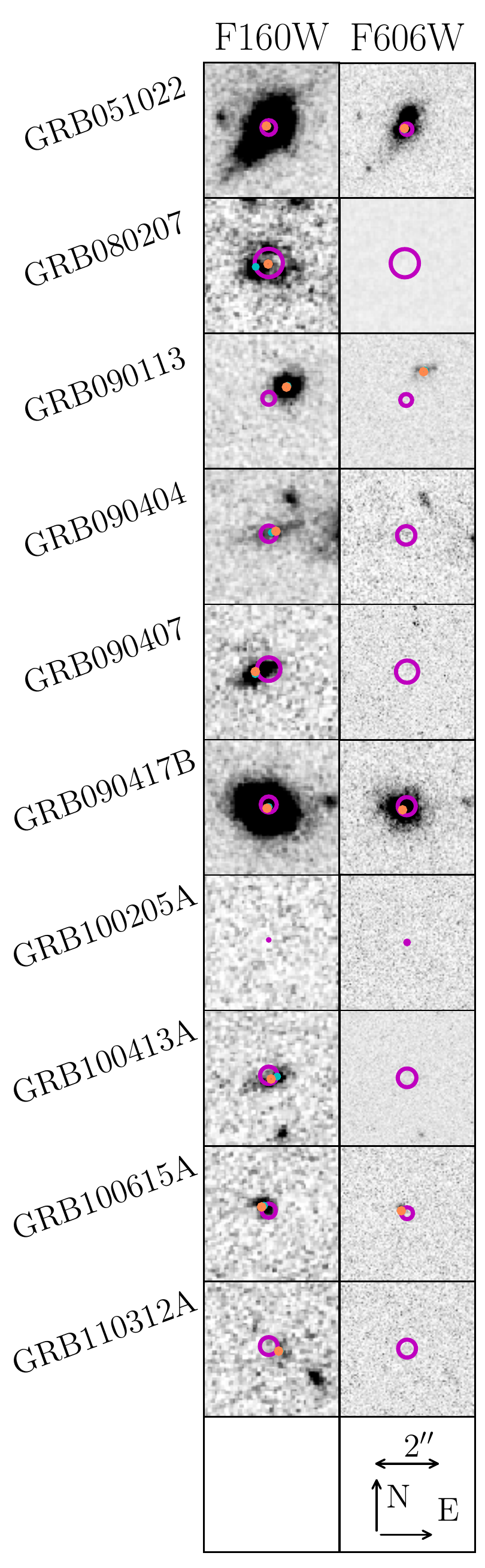}
\includegraphics[height=0.9\textheight]{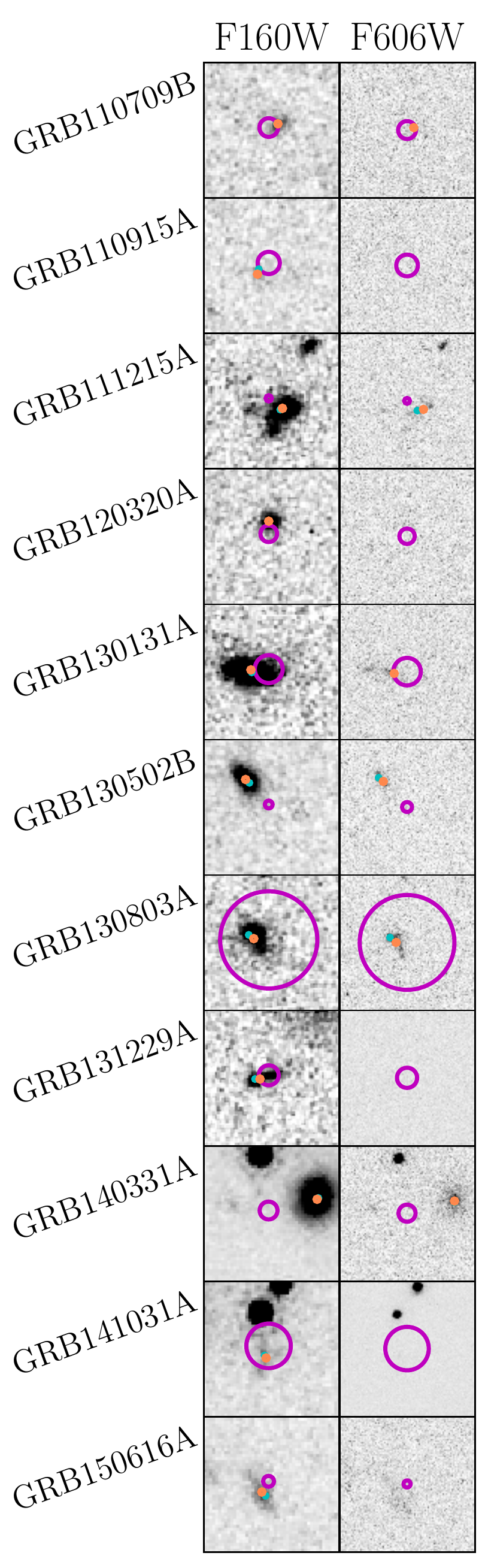}
\caption{The positions of the best available afterglow localisations on their host galaxies after astrometric alignment. Magenta circles represent the $CXO$ or other afterglow positions, and their sizes correspond to the 1${\sigma}$ confidence region. Smaller cyan and orange circles correspond to the host galaxy brightest pixel and barycentre respectively. Note that the host of GRB\,110915A is present but is largely obscured by the brightest pixel and barycentre markers.}
\label{fig:grbpositions}
\end{figure*}

Figure \ref{fig:grbpositions} shows the best position of the GRB in the {\it HST} frames for each burst. The larger, magenta error circles arise from the tie and afterglow uncertainties as described above. The smaller cyan and orange circles represent the host candidate brightest pixel and barycentre in each band respectively. 


\subsection{Host Measurement with {\sc SExtractor}}
Apparent magnitudes, enclosed flux radii and ellipticities were measured from the drizzled, charge-transfer-efficiency-corrected and filtered images with {\sc SExtractor} \citep[v2.19.5, ][]{1996A&AS..117..393B}. We resample, register and crop the IR and UVIS images using {\sc SWarp} \citep{2002ASPC..281..228B}, allowing {\sc mag{\textunderscore}auto} to be used in dual-image mode. Flux is conserved in this process with variations due to re-sampling only occurring at the millimag level.

Object identification was performed on the F160W images, using a detection threshold of at least 5 connected pixels at 1\,${\sigma}$ above the background. Non-detections are given as 3\,${\sigma}$ limits. These are calculated from 0.4 arcsec apertures in the sole case that there is no detection in F160W, using the STScI tabulated zero points\footnote{\label{note6}\url{http://www.stsci.edu/hst/wfc3/analysis/ir_phot_zpt}}. Every other non-detection is in F606W, and the aperture positions and sizes were determined in these cases by the {\sc mag{\textunderscore}auto} apertures used on the F160W image, through the use of {\sc SExtractor}'s dual-image mode. A standard smoothing filter was used on all the images, with a radius of 3 pixels. 

The cleaning parameter was also varied in order to remove spurious, spatially separated pixels which were mostly likely incorrectly attributed to a source. The appropriate zero points for each instrument, CCD and filter were obtained from the image headers and STScI\footnotemark[7]. The magnitude errors output by {\sc SExtractor} are corrected for correlated noise following \citet{2000AJ....120.2747C} and \citet{2002PASP..114..144F}. Galactic dust attenuation is calculated using the York  Extinction  Solver \citep[YES, ][]{2004AJ....128.2144M} with an $R_{V}=3.1$ Fitzpatrick reddening law \citep{1999PASP..111...63F}, effective filter wavelengths from the SVO filter profile service \citep{2012ivoa.rept.1015R,2013svo}, and the dust maps of \cite{2011ApJ...737..103S}.
Enclosed flux fraction radii measurements were performed using {\sc SExtractor} with the standard 3 pixel smoothing. Radii enclosing 20, 50 and 80 per cent of the flux were measured in each case.

\subsection{Concentration and Asymmetry}
The morphology and structure of a galaxy can provide insight into the nature of the constituent stellar populations. For example, it is well known that irregular or disturbed morphologies are associated with recent or ongoing star formation. 
Quantitative measures of galaxy morphology are provided by the concentration $C$ and asymmetry $A$ parameters \citep[][]{1985ApJS...59..115K,1996MNRAS.279L..47A,2000AJ....119.2645B,2000A&A...354L..21C,2000ApJ...529..886C,2003ApJS..147....1C,2003ApJ...596L...5C,2004AJ....128..163L}. Concentration is proportional to the log of the ratio of the radii enclosing 80 and 20 per cent of the total source flux, and measures the degree to which light is centrally concentrated within a galaxy. Asymmetry is obtained by rotating an image cutout through 180$^{\circ}$ around the barycentre of the galaxy of interest, followed by image subtraction, normalisation and summation. An identical process is carried out on blank sky regions for background asymmetry subtraction. 

We employ the same division of $CA$ parameter space as \citet{2005ApJ...633...29C}, categorising galaxies as ellipticals, spirals/irregulars, or mergers. Additionally, we use the ellipticity (i.e. one minus the ratio of semi-minor and semi-major axis length) to break the degeneracy in $CA$ parameter space between objects with similar concentrations and asymmetries but different projected 2D light distributions. This issue arises because elongated objects can be symmetric through a 180$^\circ$ rotation, and we are interested here in the effect of viewing angle and the line of sight through the GRB host. For disky galaxies, ellipticity is a proxy for line of sight inclination, information that the $CA$ parameters may not be able to provide.

We do not perform an analysis of $F_{\mathrm{light}}$ \citep{2006Natur.441..463F} or the third `CAS' parameter, clumpiness $S$. The reason for the former is that the positional uncertainties are sufficiently large that significant fractions (or in some cases, 100 per cent) of the hosts are enclosed by the error circle. \citet{2016ApJ...817..144B} showed that high error circle  to galaxy area ratios produce $F_{\mathrm{light}}$ values which are  significantly biased to lower values. The clumpiness statistic becomes increasingly unreliable as redshift increases, as demonstrated by \citet{2003ApJS..147....1C}. Pixel-to-pixel variations caused by the {\it HST} PSF create high frequency power that is not due to spatial variation in the stellar populations, which can be problematic for sources which are small in extent.   

\begin{table*}
\hspace*{-1cm}
\caption{Host properties for the entire sample of 21 GRB hosts. For brevity, the astrometric tie information shown here is for the IR images only. In most cases, tying is performed via a Pan-STARRS $r$-band intermediate (PS-$r$) as described in the text. The magnitudes have been corrected for Galactic extinction, listed in the final column, and non-detections are given as 3\,${\sigma}$ limits.} 
\begin{tabular}{l l l c c c c c c c c c c c}
\hline 
GRB & Intermediate & ${\sigma}_\mathrm{tie}$[$''$] & $m_{160}$ & ${\sigma}_{160}$ & $m_{606}$ & ${\sigma}_{606}$ & $R_\mathrm{norm}$ & $R_\mathrm{n,bp}$ & $A$ & $C$ & ${\epsilon}$ & $P_\mathrm{chance}$ & $A_V$(Gal) ${\ddagger}$\\ [0.25ex] 
\hline 
051022	&	PS-$r$	&	0.21	&	20.582	&	0.004	&	21.932	&	0.003	&	0.21	&	0.14	&	0.12	&	2.84	&	0.41	&	4.16${\times}$10$^{-5}$	& 0.150  \\
080207	&	PS-$r$	&	0.41	&	24.017	&	0.070	&	$>$26.5	&	-	&	0.06	&	0.65	&	0.17	&	1.98	&	0.27	&	4.96${\times}$10$^{-5}$	& 0.057 \\
090113	&	PS-$r$	&	0.19	&	22.705	&	0.022	&	24.225	&	0.017	&	2.25	&	2.41	&	0.16	&	2.57	&	0.16	&	1.27${\times}$10$^{-2}$	& 0.205  \\
090404	&	PS-$r$	&	0.24	&	23.334	&	0.043	&	$>$25.6	&	-	&	0.39	&	0.17	&	0.07	&	2.69	&	0.7	&	2.32${\times}$10$^{-3}$	& 0.051  \\
090407	&	PS-$r$	&	0.34	&	22.779	&	0.037	&	$>$26.7	&	-	&	1.23	&	1.29	&	0.2	&	2.71	&	0.5	&	4.35${\times}$10$^{-3}$	& 0.168  \\
090417B	&	PS-$r$	&	0.23	&	20.595	&	0.004	&	21.425	&	0.003	&	0.34	&	0.19	&	0.12	&	3.31	&	0.27	&	8.92${\times}$10$^{-5}$	& 0.041  \\
100205A	&	Direct GN	&	0.02	&	$>$26.7	&	-	&	$>$27.1	&	-	&	 -	&	 -	&	-	&	-	&	-	&	 -	& 0.047  \\
100413A	&	PS-$r$	&	0.25	&	23.667	&	0.077	&	25.947	&	0.134	&	0.42	&	0.83	&	0.16	&	2.33	&	0.03	&	7.38${\times}$10$^{-4}$	& 0.281  \\
100615A	&	PS-$r$	&	0.20	&	23.912	&	0.058	&	24.972	&	0.041	&	1.41	&	1.19	&	0.22	&	2.71	&	0.19	&	2.10${\times}$10$^{-3}$	& 0.111  \\
110312A	&	PS-$r$	&	0.26 ${\dagger}$	&	24.806	&	0.204	&	$>$26.8	&	-	&	1.05	&	1.15	&	0.01	&	2.26	&	0.21	&	6.58${\times}$10$^{-3}$	& 0.095  \\
110709B	&	PS-$r$	&	0.27	&	24.829	&	0.010	&	26.549	&	0.181	&	1.25	&	1.12	&	0.19	&	2.71	&	0.18	&	6.15${\times}$10$^{-3}$	& 0.121  \\
110915A	&	PS-$r$	&	0.31	&	25.628	&	0.171	&	$>$27.5	&	-	&	1.92	&	1.42	&	0.23	&	2.61	&	0.24	&	2.12${\times}$10$^{-2}$	& 0.142  \\
111215A	&	PS-$r$	&	0.09	&	22.361	&	0.032	&	24.071	&	0.035	&	1.72	&	1.6	&	0.19	&	2.7	&	0.42	&	5.99${\times}$10$^{-3}$	& 0.156 \\
120320A	&	PS-$r$	&	0.24	&	23.940	&	0.069	&	$>$27.0	&	-	&	1.42	&	1.42	&	0.04	&	2.5	&	0.01	&	4.63${\times}$10$^{-3}$	& 0.073  \\
130131A	&	Direct UKIRT	&	0.40	&	21.889	&	0.022	&	24.089	&	0.037	&	1.85	&	1.77	&	0.15	&	2.84	&	0.51	&	8.53${\times}$10$^{-3}$	& 0.038  \\
130502B	&	PS-$r$	&	0.11	&	22.612	&	0.026	&	24.642	&	0.026	&	6.93	&	6.93	&	0.1	&	2.7	&	0.22	&	1.52${\times}$10$^{-1}$	& 0.515  \\
130803A	&	XRT only	&	1.40	&	22.740	&	0.037	&	23.730	&	0.026	&	 -	&	 -	&	0.19	&	2.69	&	0.21	&	 -	& 0.140  \\
131229A	&	PS-$r$	&	0.29 ${\dagger}$	&	23.235	&	0.077	&	$>$25.8	&	-	&	0.87	&	1.31	&	0.19	&	2.41	&	0.69	&	3.24${\times}$10$^{-3}$	& 0.671  \\
140331A	&	PS-$r$	&	0.26	&	20.127	&	0.007	&	23.127	&	0.022	&	3.29	&	3.39	&	0.08	&	3.31	&	0.08	&	7.32${\times}$10$^{-3}$	& 0.112  \\
141031A	&	PS-$r$	&	0.64 ${\dagger}$	&	22.812	&	0.032	&	$>$25.7	&	-	&	 0.94	&	 0.89	&	0.19	&	2.33	&	0.32	&	 4.47${\times}$10$^{-3}$	& 0.423  \\
150616A	&	FORS2	&	0.16	&	22.870	&	0.051	&	24.250	&	0.051	&	1.11	&	1.09	&	0.2	&	2.43	&	0.47	&	5.42${\times}$10$^{-3}$	& 0.093  \\
\hline 
\end{tabular}
\newline
 ${\dagger}$ - These hosts have a barycentre uncertainty of more than 0.1\,arcsec. ${\ddagger}$ - This is the F606W band Milky Way extinction.
\label{tab:noz}
\end{table*}

\subsection{Morphological Uncertainties}
Due to the drizzling process, there is correlated noise in the final {\it HST} images which is not accounted for in the [{\sc ERR}] maps output by the data reduction. To address this issue, we resampled the pixels in the pre-drizzled {\sc flc} and {\sc flt} images by adding values sampled from their {\sc [ERR]} extension uncertainty distribution, {\it before} drizzling. This process was repeated a few hundred times for each set of images, with measurements made on the new drizzled image each time. This produced distributions of {\sc SExtractor} and $CA$ output parameters \citep[following the methodology of ][]{2017MNRAS.467.1795L}. Our $CA$ results use the mean of the drizzled image measurements, with uncertainties given by the 1\,${\sigma}$ spread of the re-sampled distribution. In cases where the original, observed measurement falls outside this region, we use it as the upper or lower limit as appropriate. The quantities for which uncertainties are estimated in this way are the enclosed flux radii, host barycentre, asymmetry and ellipticity.

\begin{table*}
\centering 
\caption{GRB host properties for which a redshift is required. Uncertainties are given on the redshift of GRBs\,140331A, as this is not spectroscopically determined.} 
\begin{tabular}{l c l c c c c c c c c c c }
\hline 
GRB & $z$ & $z$ ref & Scale & $R_{20}$ & $R_{50}$ & $R_{80}$ & R$_\mathrm{phys}$ & $M_{160}$ & $M_{606}$ \\ [0.25ex] 
\newline
 &  & & [kpc/$''$] & [kpc] & [kpc] & [kpc] & [kpc] &  & \\ 
\hline 
051022	&	0.809	&				[1]		&	7.536	&	1.48	&	2.94	&	5.47	&	0.63	&	-22.177$^{+0.004}_{-0.004}$	&	-20.958$^{+0.003}_{-0.003}$	\\
080207	&	2.0858	&				[2]		&	8.328	&	3.21	&	5.09	&	7.98	&	0.30	&	-20.778$^{+0.070}_{-0.070}$	&	$>$-18.4	\\
090113	&	1.7493	&				[2]		&	8.456	&	1.35	&	2.60	&	4.41	&	5.86	&	-21.613$^{+0.022}_{-0.022}$	&	-20.279$^{+0.017}_{-0.017}$	\\
090407	&	1.4485	&				[2]		&	8.449	&	1.39	&	2.77	&	4.84	&	3.40	&	-21.196$^{+0.037}_{-0.037}$	&	$>$-17.4	\\
090417B	&	0.345	&				[3]		&	4.894	&	0.87	&	1.87	&	3.61	&	0.64	&	-20.355$^{+0.004}_{-0.004}$	&	-19.562$^{+0.003}_{-0.003}$	\\
100615A	&	1.398	&				[4]		&	8.431	&	0.71	&	1.45	&	2.49	&	2.04	&	-20.042$^{+0.058}_{-0.058}$	&	-19.080$^{+0.041}_{-0.041}$	\\
110709B	&	2.109	&				[5] $^{a}$	&	8.315	&	0.94	&	2.04	&	3.28	&	2.56	&	-19.930$^{+0.010}_{-0.010}$	&	-19.419$^{+0.045}_{-0.045}$	\\
111215A	&	2.012	&				[6] $^{b}$	&	8.364	&	1.31	&	2.67	&	4.54	&	4.60	&	-22.275$^{+0.032}_{-0.032}$	&	-20.706$^{+0.035}_{-0.035}$	\\
140331A	&	1.00$^{+0.11}_{-0.04}$		&	[7] $^{c}$	&	8.008	&	1.51	&	3.51	&	6.97	&	11.55	&	-23.120$^{+0.222}_{-0.088}$	&	-20.222$^{+0.223}_{-0.090}$	\\
\hline 
\end{tabular}
\newline
[1] - \citet{2007A&A...475..101C}, [2] - \citet{2012ApJ...758...46K}, [3] - \citet{2009GCN..9156....1B}, [4] - \citet{2013GCN.14264....1K}, [5] - \citet{2016ApJ...817....7P}, [6] - \citet{2015MNRAS.446.4116V} , [7] - \citet{2018MNRAS.478....2C}
\newline
$a$ - this is a tentative redshift based on one emission line, $b$ - tentative and based on a single line, but consistent with the photometric redshift, $c$ - this is a photometric redshift
\label{tab:needsz}
\end{table*}


\begin{table}
\centering 
\caption{Here, we provide upper limits on the physical separations, $R_\mathrm{phys}$, and enclosed flux radii for those GRBs without redshifts. A redshift of ${\sim}$1.6 is assumed, which corresponds to the maximum angular diameter distance and (approximately) the mean GRB redshift. We note that GRB\,100413A has an XRT-derived redshift of $z {\sim}$ 4 \citep{2010GCN.10588....1C}, but a lower $z$ solution cannot be ruled out.} 
\begin{tabular}{l c c c c}
\hline 
GRB & $R_\mathrm{phys}$ & $R_{20}$ & $R_{50}$ & $R_{80}$ \\ [0.25ex] 
\newline
 & [kpc] & [kpc] & [kpc] & [kpc] \\ [0.25ex] 
\hline 
090404	&	2.22	&	2.76	&	5.71	&	9.54	\\
100413A	&	1.10	&	1.42	&	2.47	&	4.14	\\
110312A	&	2.73	&	1.4	&	2.6	&	3.97	\\
110709B	&	2.61	&	0.96	&	2.08	&	3.34	\\
110915A	&	3.88	&	1.09	&	2.03	&	3.63	\\
120320A	&	3.05	&	1.14	&	2.14	&	3.61	\\
130131A	&	6.52	&	1.74	&	3.53	&	6.45	\\
130502B	&	18.44	&	1.34	&	2.66	&	4.66	\\
130803A	&	3.68	&	1.7	&	3.13	&	5.87	\\
131229A	&	2.32	&	1.42	&	2.66	&	4.31	\\
141031A	&	6.67	&	0.77	&	3.39	&	5.07	\\
150616A	&	3.8	&	1.84	&	3.44	&	5.64	\\
\hline 
\end{tabular}
\label{tab:damax}
\end{table}

\section{Results}\label{sec:results}
Basic results, as measured from the photometric images are reported in tables \ref{tab:noz} and \ref{tab:needsz}. We use redshifts where they are known to calculate angular and luminosity distances, providing physical scales and absolute magnitudes. Where there is no redshift information, an upper limit on physical offsets and sizes can still be obtained since the angular diameter distance reaches a maximum at $z{\sim}$1.6, for the cosmological parameters which we have adopted. This also happens to be similar to the mean redshift of GRBs, and there is little 
evolution in the the angular diameter over the redshift range $1<z<3$. These limits are listed in table \ref{tab:damax}.

In order to judge whether an association between a burst and the nearest galaxy is genuine, we use the $H$-band galaxy counts of \citet{2006MNRAS.370.1257M} and perform a false alarm probability analysis \citep[e.g.,][]{2002AJ....123.1111B,2007MNRAS.378.1439L}. The probability $P_{\mathrm{ch}}$ of finding at least one unrelated source of magnitude $m_{g}$ or brighter within an angular separation $r$ is given by, 
\begin{equation}
P_{\mathrm{ch}} = 1 - e^{-{\Sigma}(m{\leq}m_{g}){\pi}r^{2}},
\end{equation}
where ${\Sigma}$ is the surface density of sources. Using a cut-off of $P_{\mathrm{ch}}<0.05$, which permits at most one interloper in our sample, we reject 1 candidate (GRB\,130502B) as a potential chance alignment based on this cut. All further analyses therefore omit GRB\,130502B. If we consider this to be a change alignment, the implication is that the true host is undetected, and possibly very distant. We note, however, that the X-ray brightness of this burst makes a high redshift scenario unlikely \citep{2009MNRAS.397.1177E,2013GCN.14540....1M}. While we do not include the burst and its candidate host in this sample, we acknowledge the possibility that the nearby galaxy is associated but that the burst is at a large offset. 
The probability of there being 1, 2 and 3 interlopers in the remainder of the sample (ignoring GRB\,130803A whose positional uncertainty is too large for a meaningful $P_{\mathrm{ch}}$ estimate) is calculated using the Poisson Binomial distribution, giving 0.1, 3.8${\times}$10$^{-3}$ and 9.8${\times}$10$^{-5}$ respectively \citep{Hong:2013:CDF:2400772.2401582}.

%

\subsection{Host Colours and Luminosities}\label{sec:colmag}

\begin{figure}
\centering
\includegraphics[width=0.9\columnwidth]{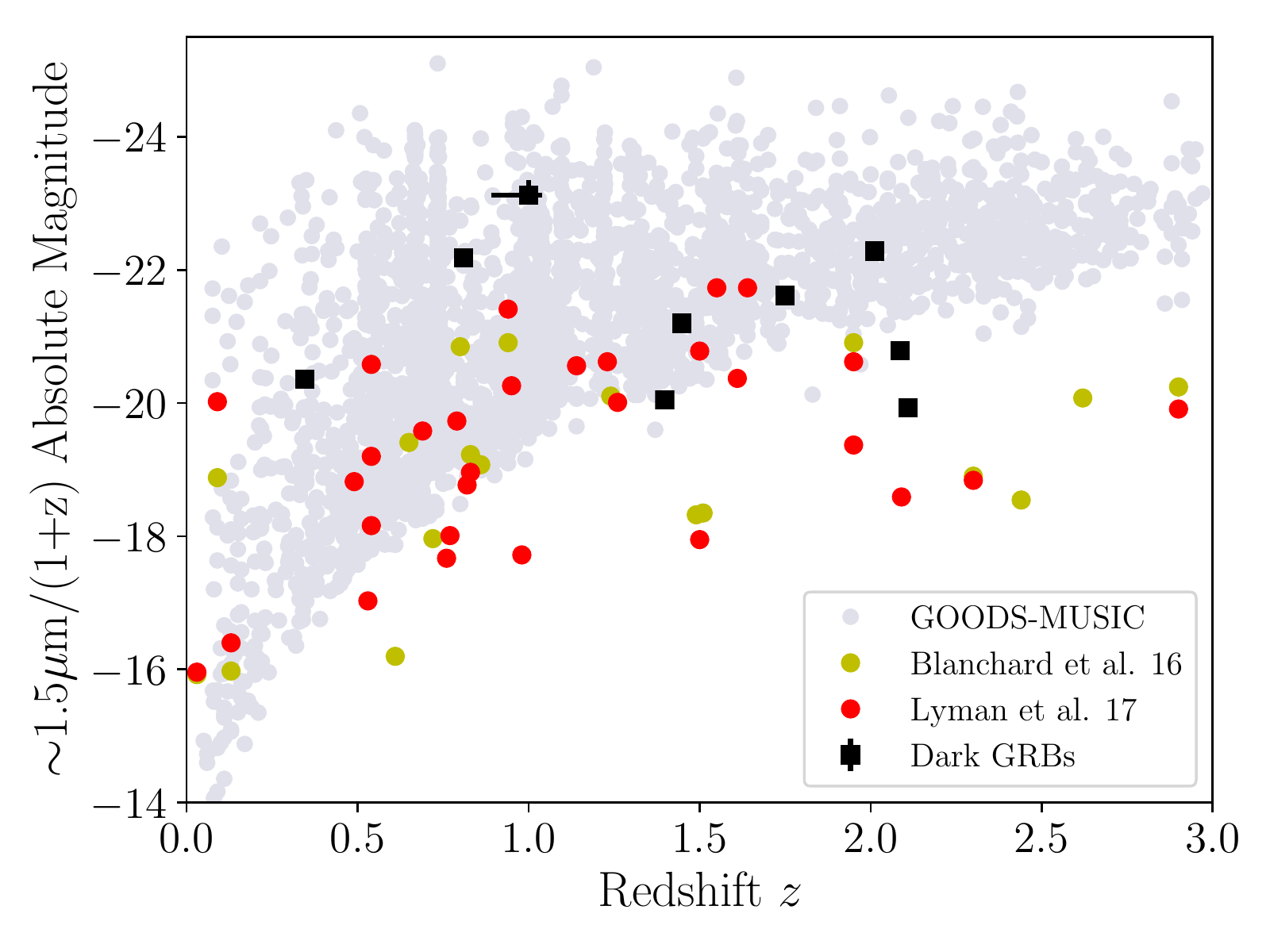}
\caption{Redshift versus absolute magnitude for the dark GRB hosts, comparison galaxies hosting optically bright GRBs, and the GOODS-MUSIC galaxy catalogue. The hosts of dark GRBs are more luminous than bright GRB hosts over a wide range in redshift.}
\label{fig:zmags}
\end{figure}

We use the F160W absolute magnitudes and F606W-F160W colours to further characterise the host population. Figure \ref{fig:zmags} compares a proxy for the absolute magnitudes of the hosts in this sample, calculated as $m_{F160W}$ - ${\mu}$ + 2.5log$_{10}$(1+$z$) (where ${\mu}$ is the distance modulus), to those from \citet{2016ApJ...817..144B} and \citet{2017MNRAS.467.1795L}. For all our results, where there is overlap between samples, our measurements agree well \citep[e.g.][]{2015MNRAS.446.4116V,2016ApJ...817..144B}. The dark hosts are more luminous than the hosts of optically bright GRBs, at all redshifts. They are also more representative of the general star forming galaxy population at their epoch, where we show all galaxies reported by \citet{2006A&A...449..951G} and \citet[][labelled GOODS-MUSIC]{2009yCat..35040751S}. Previous GRB hosts studies have found similar results \citep[e.g.,][]{2012ApJ...756..187H, 2013ApJ...778..128P}, and attribute dark GRB host luminosities to greater stellar masses, consistent with a dust-extinguished afterglow scenario. 

\begin{figure}
\centering
\includegraphics[width=0.95\columnwidth]{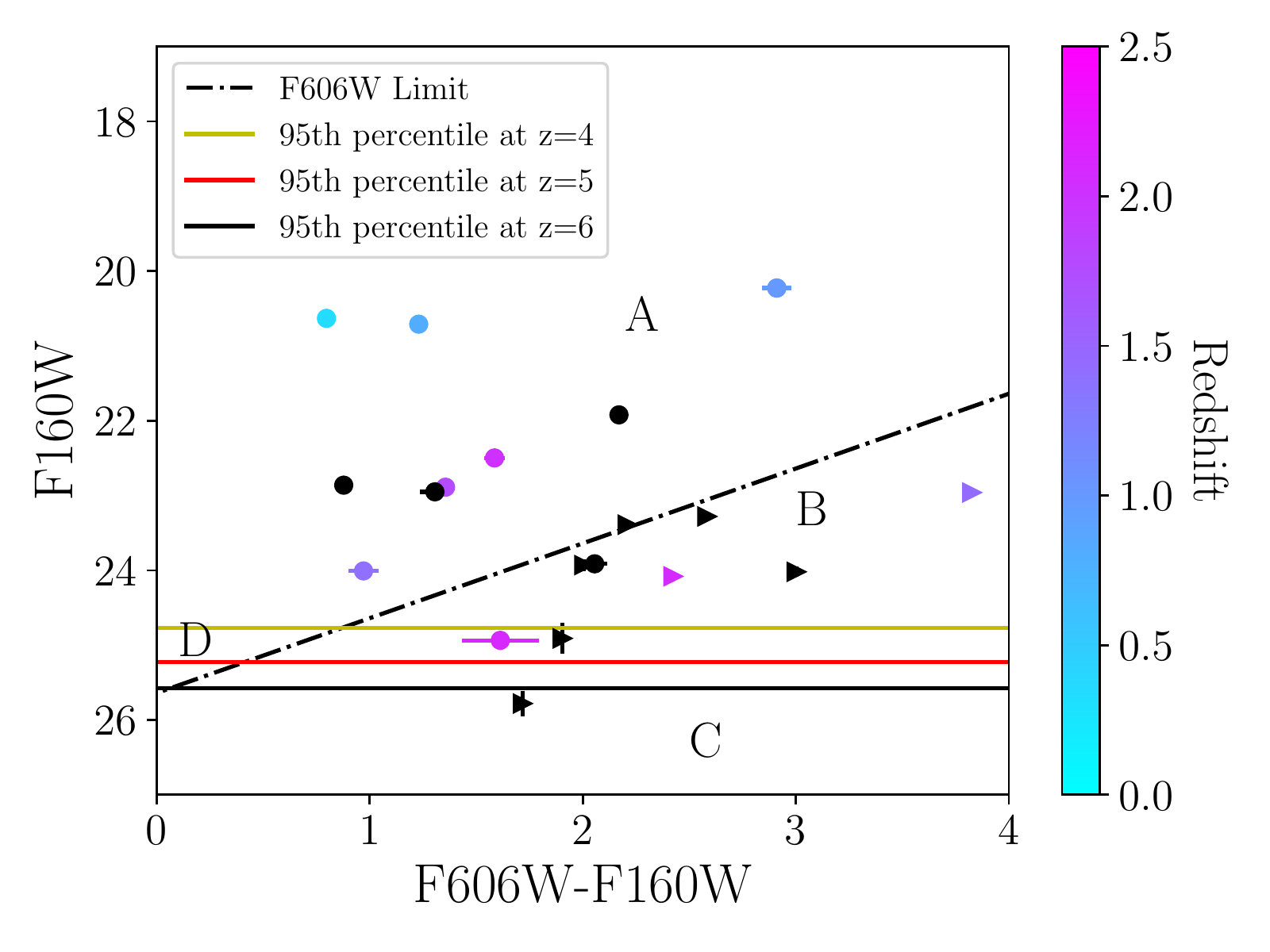}
\caption{The F160W apparent magnitude of GRB hosts in this sample versus their observed F606W-F160W colour. Points which are coloured have a redshift, those in black do not. Horizontal lines denote  the apparent UV magnitude fainter than which 95 per cent of the UV star-forming galaxy population would be observed \citep{2015ApJ...803...34B}. 
The dashed line represents the shallowest F606W limit in our sample. Sources to the left of the dashed line, in regions A and D, are detected in F606W and a high redshift scenario is disfavoured. In region B, there are sources which are undetected in F606W but which are implausibly bright in F160W if they lie at $z>5$. The two undetected sources in region C (depending on the redshift) are also faint in F160W, and are possible high $z$ candidates. One source, GRB\,100205A does not appear as it was not-detected in either band.}
\label{fig:colours}
\end{figure}

The colours of these hosts also provide information. In figure \ref{fig:colours}, we show a colour-magnitude diagram plotting apparent F160W magnitudes against F606W-F160W colour. The points themselves are coloured according to redshift, if available, otherwise they are left black. Horizontal lines denote constraints derived from the luminosity function of high redshift galaxies reported by \citet{2015ApJ...803...34B}. At each of $z=4, 5$ and 6, we use the reported M$^\ast$ and faint end slope to determine the apparent (rest-frame UV) magnitude fainter than which 95 per cent of the UV star-forming galaxy population would be observed. At $z>4$, the observed $H$-band lies below the Lyman break, and for star forming galaxies, the mean spectral energy distribution is approximately flat in $F_{\nu}$ and therefore AB magnitude. This leads to a zero colour term between the rest-UV and rest-optical (observed $H$-band). We assume that the luminosity function extends down to an absolute magnitude $M_\mathrm{UV}=-10$. We note that the line shown at $z\sim5$ is within a few tenths of a magnitude of the brightest galaxy detected at $z>5$ in the GOODS-MUSIC moderately wide area photometric survey \citep{2006A&A...449..951G,2009yCat..35040751S}.

We use the set of 95$^{\mathrm{th}}$ percentile limits discussed above, and the shallowest F606W limit in our imaging, to split this parameter space into 4 (redshift-dependent) regions. Objects in region A lie above both limits. They would have to be exceptionally bright if at $z>5$, and are in fact detected in F606W suggesting that the Lyman break does not lie red-wards of the F606W band, which is centred around ${\sim}$5700{\AA} \citep{2012ivoa.rept.1015R}. Therefore, we can say with confidence that all sources in region A are at $z<4$, providing a limit for three galaxies without previous redshift information: GRBs\,130131A, 130803A and 150616A. In region B of figure \ref{fig:colours}, the lower bound of which is redshift dependent, the galaxies would have to be unusually bright in F160W to be at high redshift but are nonetheless undetected or very faint in F606W. Such sources are most likely intrinsically red. Sources in region C, below both limits, are either undetected or faint in F606W, and faint enough in F160W to plausibly be at $z>4$ (or $z>5$ or $z>6$ depending on the adopted limit). While the F606W non-detections here could be attributed to the Lyman break, a faint and dusty scenario cannot be ruled out. Two sources lie clearly in region C. One of these, the  non-detection of GRB\,100205A in either band, is not shown. This burst has previously been suggested as a high redshift candidate \citep{2010GCN.10374....1C}. The other is GRB\,110915A. 
An additional two objects lie in an ambiguous region, on the boundary between regions B and C at $z=4-5$, and would have to be luminous if at high redshift, although not exceptionally so. Of these, one is known to lie at $z\sim2$.
Finally, objects in region D would be detected in F606W and are thus very likely at $z<4$.

\subsection{Host Morphologies}
\begin{figure*}
\centering
\includegraphics[width=0.95\textwidth]{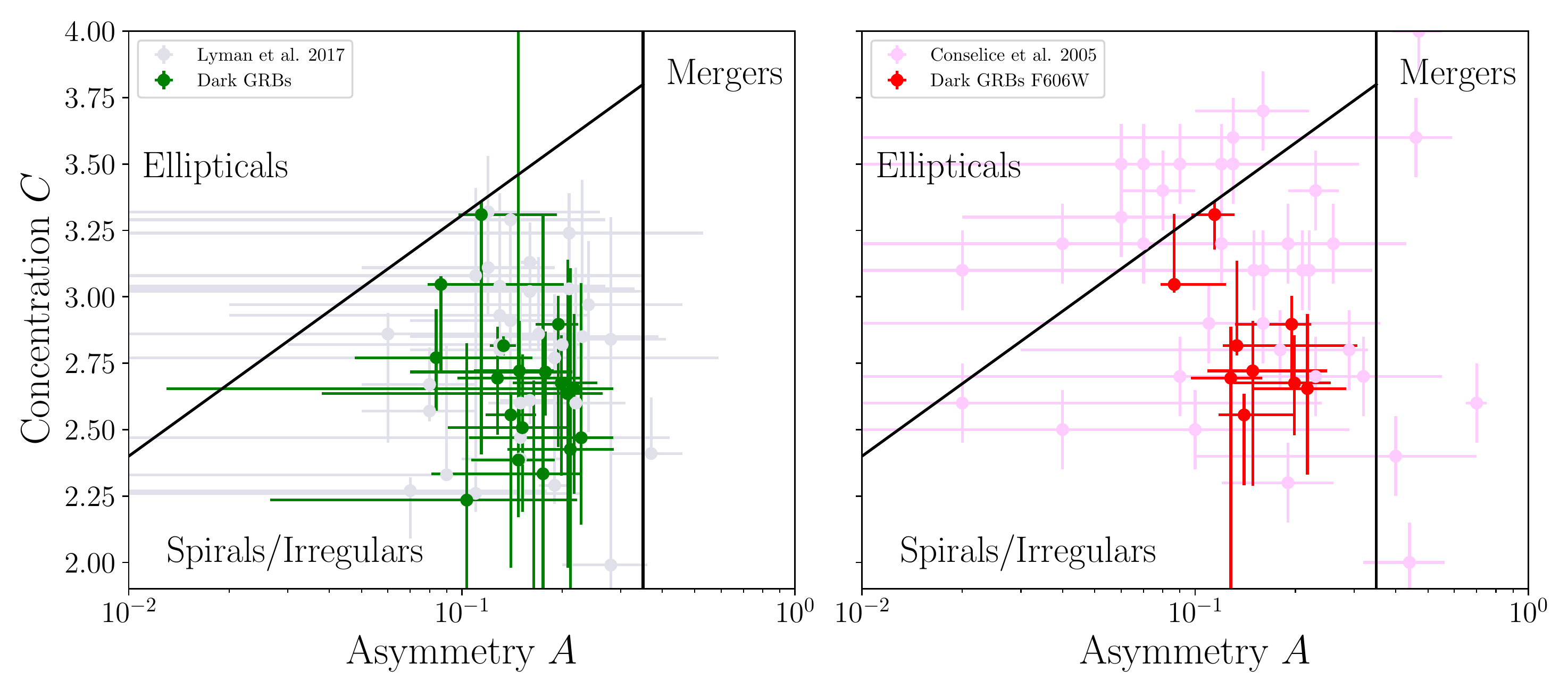}
\caption{The position of the 19 F160W-detected dark GRB hosts in $CA$ parameter space (left), and the 9 detected in F606W for which measurements could be made (right). Comparisons are made to \citet[shown in grey]{2017MNRAS.467.1795L} and \citet[pink]{2005ApJ...633...29C}. The most significant difference occurs for concentration in the F606W band.}
\label{fig:f160w_CA}
\end{figure*}

Morphological measurements for each detected galaxy are also listed in table \ref{tab:noz}. Figure \ref{fig:f160w_CA} shows the $CA$ results for the dark hosts in the F160W band, with equivalent measurements in the same band from \citet{2017MNRAS.467.1795L}  shown in grey. The dark population shows no statistically significant offset from the normal bursts. We also measure $CA$ parameters in the F606W band. In two cases where there is a photometric detection, the galaxy appears sufficiently diffuse in F606W that meaningful $CA$ measurements could not be made (i.e. the source is frequently undetected as an object when re-drizzling is performed, this occurred for GRBs\,100413A and 150616A). The true noise in the image is given by the pixel-to-pixel noise mutilplied by a corrective factor (order of magnitude ${\sim}$10) that accounts for correlated noise from drizzling \citep{2002PASP..114..144F}. Due to this correlated noise, re-sampling produces more variation in the resultant re-drizzled galaxy pixel values and explains the large morphological uncertainties that are obtained with this method.

The F606W results are compared to the optical measurements of \citet{2005ApJ...633...29C}. An Anderson-Darling (AD) test produces a p-value of 0.048 for concentration (providing marginal evidence that dark GRB hosts are more concentrated) and 0.25 (i.e. no significant difference) for asymmetry. 


If dark GRB hosts are preferentially edge-on and disky, the average line-of-sight through the galaxy to the burst location would be longer and more prone to dust extinction in the plane of the disk. This would manifest as dark GRB hosts showing relatively elongated morphologies which are identified by the ellipticity ${\epsilon}$ but not necessarily the asymmetry. Again, statistical comparison between this sample and that of \citet{2017MNRAS.467.1795L} provides no grounds to reject the null hypothesis that the two populations have the same underlying ellipticity distribution. Comparing the physical half-light radius $R_{50}$ of the dark GRB hosts with a known redshift from table \ref{tab:needsz} to the samples of \citet{2017MNRAS.467.1795L} and \citet{2016ApJ...817..144B}, we find that the dark GRB hosts are physically more extended, with AD tests yielding p-values of 0.030 and 0.087 respectively. The median physical $R_{50}$ of this sample is 2.7${\pm}$0.4\,kpc, versus 1.8${\pm}$0.1\,kpc for \citet{2016ApJ...817..144B} and 1.7${\pm}$0.2\,kpc for \citet{2017MNRAS.467.1795L}.

\subsection{Burst Offsets}\label{sec:burstoffsets}
\begin{figure*}
  \centering
  \begin{minipage}[b]{0.95\textwidth}
    \hspace*{0.02\textwidth}
    \includegraphics[width=0.95\textwidth]{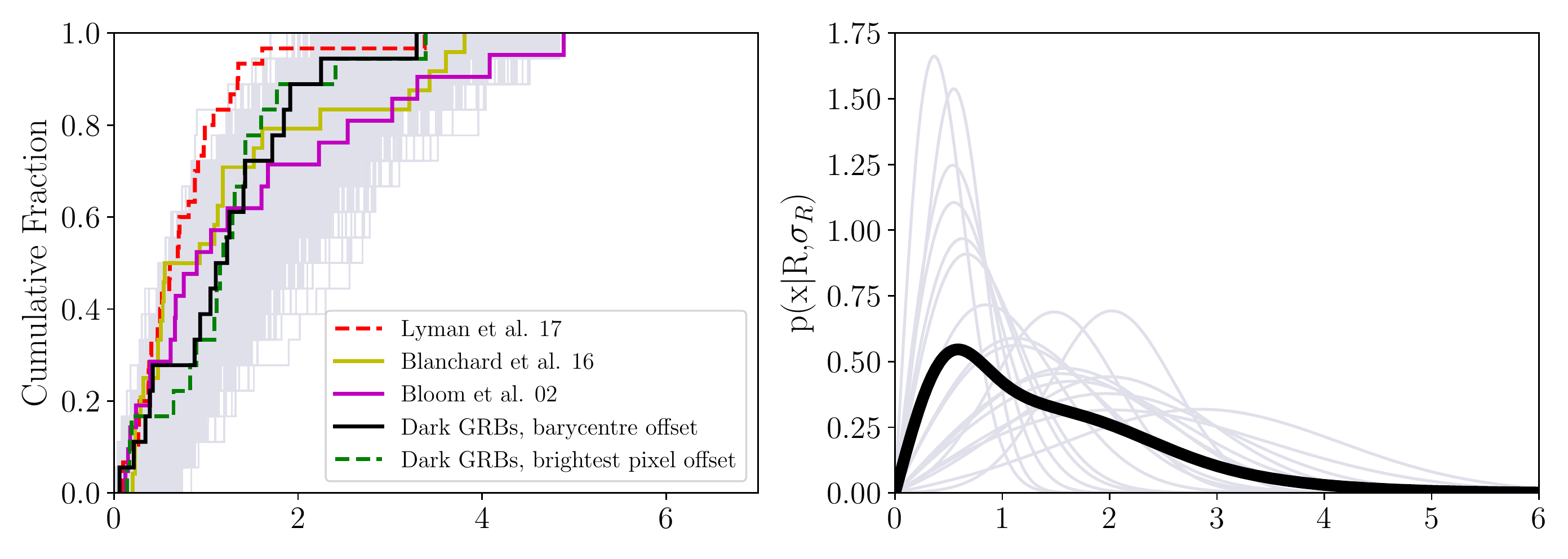}
  \end{minipage}
  \vfill
  \begin{minipage}[b]{0.95\textwidth}
    \hspace*{0.02\textwidth}
    \includegraphics[width=0.95\textwidth]{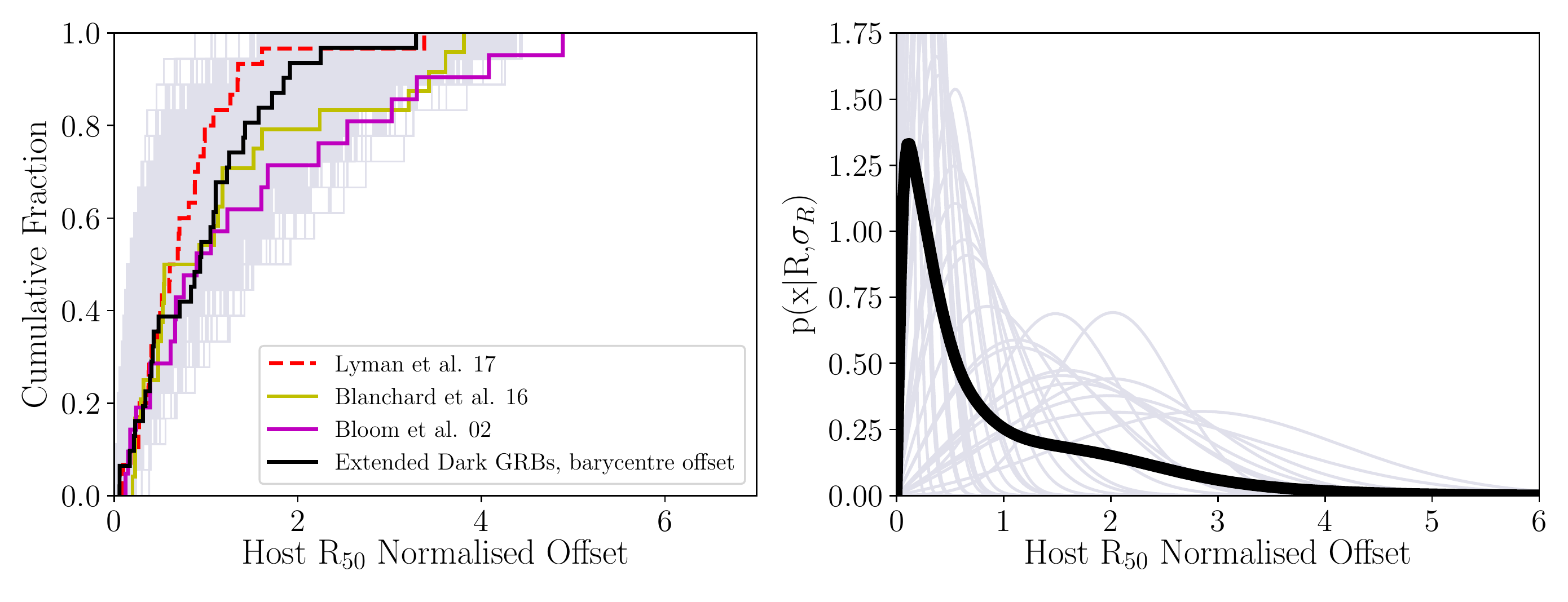}
    \caption{Upper left: The cumulative distribution of R$_{50}$ normalised host offset for the dark GRBs and a selection of comparison data sets. The grey distributions in the background are Monte Carlo resamples, drawn from the summed Ricean distribution of the dark sample. Upper right: The individual Ricean probability distributions for the dark GRB normalised offsets are shown in grey, with the summed and normalised distribution overlaid in black. Lower panels: As above, but for the extended dark burst sample, including optically or NIR detected dark GRBs from \citet{2016ApJ...817..144B}.}
    \label{fig:f160w_offsets}
  \end{minipage}
\end{figure*}

Out of the 21 bursts in our sample, one host candidate is rejected as a potential chance alignment, one host is undetected and one GRB has a positional uncertainty which is too large to measure a meaningful offset from the putative host. We therefore measure the offset of the burst from the host light barycentre ($R_\mathrm{norm}$) and brightest pixel ($R_\mathrm{n,bp}$) for the remaining 18 GRBs. The offset distribution, normalised by the host $R_{50}$ radius, is shown for the F160W band in the upper left panel of figure \ref{fig:f160w_offsets}. This is compared to distributions drawn from the literature. The uncertainty on each offset has contributions from afterglow, tie and host positional uncertainties. In order to quantify the uncertainty on the cumulative distribution, we follow the approach of \citet{2002AJ....123.1111B} and \citet{2016ApJ...817..144B}. In the case of a point source with approximately Gaussian uncertainties, offset from the host centre by a distance $R$, the frequency of occurrence at any given offset $x$ is described by the Rice distribution. The upper right panel of figure \ref{fig:f160w_offsets} shows the Ricean probability density functions for each GRB in our sample individually, and the summed distribution. We randomly draw 18 barycentre offsets from the summed distribution 1000 times, and plot these in grey in the left-hand panel. This gives an indication of the uncertainty on the cumulative distribution. 

\begin{table}
\centering 
\caption{Statistical tests comparing the dark GRB host-normalised barycentre offset distribution to other samples, using bursts solely from this sample (upper set of statistics) and including optically/NIR detected dark bursts from \citet{2016ApJ...817..144B} (lower set). The first column is the significance of the difference between the median $R_\mathrm{norm}$ values. The second is the frequency with which randomly drawn offsets are less than the median of the comparison sample. The final column lists AD test results.} 
\begin{tabular}{l c c c c}
\hline 
Sample & R$_\mathrm{norm}$ & ${\sigma}$ & Bootstrap & AD Test \\ [0.25ex] 
\newline
 & &  & \% & p-val \\ [0.25ex] 
\hline 
This work & 1.2${\pm}$0.3 \\
\citet{2002AJ....123.1111B}	&	0.8${\pm}$0.3 & 0.83 & 6.2	&	0.58 \\
\citet{2016ApJ...817..144B} &	0.7${\pm}$0.2 & 1.24 & 2.3	&	0.56  \\
\citet{2017MNRAS.467.1795L} &	0.6${\pm}$0.1 & 1.70 & 0.5	&	0.03  \\
\hline 
Extended sample$^{\dagger}$ & 0.9${\pm}$0.2 \\
\citet{2002AJ....123.1111B}	&	0.8${\pm}$0.3 & 0.35 & 83.9	&	0.34 \\
\citet{2016ApJ...817..144B} &	0.7${\pm}$0.2 & 0.70 & 74.8	&	0.57  \\
\citet{2017MNRAS.467.1795L} &	0.6${\pm}$0.1 & 1.17 & 61.7	&	0.20  \\
\hline 
\end{tabular}
 ${\dagger}$ - Includes these additional 13 GRBs: 050401, 060719, 061222A, 070306, 070508, 080325, 080605, 080607, 081109A, 081221, 090709A, 100621A, 120119A
\label{tab:offsetstats}
\end{table}

Table \ref{tab:offsetstats} shows the results of statistical comparisons between the dark GRB, host-normalised offset distribution reported here, and other literature samples. 
Statistical values indicating consistency between samples with a probability of $<5$ per cent are shown in bold. The results vary from being statistically consistent to marginally inconsistent. Working instead with physical offsets produces comparable results. Interestingly, the dataset most inconsistent is the normal GRB sample of \citet{2017MNRAS.467.1795L}.

Because the dark burst sample size is modest, and {\it Chandra} determined positions typically have larger positional uncertainties, we consider the effect of including additional optically or NIR detected dark GRBs in the offset comparisons. Specifically, we select dark bursts from \citet{2016ApJ...817..144B}, whose data reduction and analysis is similar to the methods we have employed. We choose bursts which have F160W (or similar) {\it HST} imaging, a ${\beta}_{\mathrm{OX}} < 0.5$ \citep[see][]{2016ApJ...817....7P}, a Galactic A$_{V}$ < 0.5 and an optical or NIR afterglow detection. The 13 host normalised offsets of those GRBs reported in \citet{2016ApJ...817..144B} which meet these criteria are added to the sample described in this paper, to create an extended dark burst offset sample. The same statistical comparisons are made between the extended sample, and the three literature datasets. The effect of including these optically or NIR detected dark bursts is shown in the bottom two panels of \ref{fig:f160w_offsets}. We find that the extended sample is also consistent with the comparison samples.

\section{Discussion}\label{sec:discuss}
\subsection{Host Colours and Magnitudes} 
In common with previous work \citep{2011A&A...526A..30G,2012ApJ...756..187H,2013ApJ...778..128P},  we find that
dark GRB hosts are typically more luminous, and thus likely of higher stellar mass, than the hosts of GRBs with bright afterglows at the same rest-frame wavelength. They also show larger physical sizes. For these reasons, the darkness of most GRBs is now widely attributed to dust within the host. \citet{2016ApJ...817....7P} estimate that no more than 2 per cent of GRBs with fluence greater than 10$^{-6}$\,erg\,cm$^{-2}$ lie at $z>5.5$. Based on the colours and magnitudes of our sample, we find that the fraction of dark GRBs in this sample which could feasibly be high redshift is 0.14${\pm}$0.08. Allowing for poisson uncertainties on the small number statistics, and assuming that 20 per cent of all GRBs are dark \citep{2011A&A...526A..30G,2012ApJ...752...62J,2016ApJ...817....7P}, this puts $<4.4$ per cent of all GRBs at $z>5$, in good agreement with previous estimates \citep[$2-3$ per cent,][]{2009AJ....138.1690P,2016ApJ...817....7P}. This is strictly an upper limit, as the non-detection of some of our targets as shown in figure \ref{fig:colours} might also be the result of moderate extinction at intermediate redshifts, particularly since the F606W band is increasingly affected by dust extinction as it probes further into the NUV.

\subsection{Host Galaxy Morphologies}
We find marginal evidence for dark GRB hosts being more concentrated in the F606W band, and no evidence for differences in asymmetry or ellipticity between our sample and those of \citet{2017MNRAS.467.1795L} and \citet{2005ApJ...633...29C}. The ellipticity result implies that the dark GRB hosts are typically not edge-on disks. In both bands, we are comparing with a similarly observed and analysed dataset. 

We might have expected that galaxies would appear to be more asymmetric in F606W, given that this corresponds to the the rest frame UV at redshift ${\sim}$ 2, and the irregularity of star forming clumps might be measurable. However, previous $CA$ analyses have shown that this effect manifests itself primarily in the clumpiness parameter \citep{2013ApJ...774...47L}, with $A$ and $C$ unaffected. The dark host sample has very similar mean $C {\sim}$ 2.6 and $A {\sim}$ 0.15
values to the $H$-band, $M_{\star}$ > 10$^{9}$\,$M_\odot$, spectroscopically-selected star forming galaxy sample of \citet{2013ApJ...774...47L}.

The most striking difference between our sample and those of \citet{2017MNRAS.467.1795L} and \citet{2005ApJ...633...29C} is that the \citet{2005ApJ...633...29C} sample is significantly more spread out over $CA$ parameter space in F606W, with mergers and ellipticals clearly identified. There is also a marginal trend for the dark hosts to be more concentrated in this band, as the AD test p-value of 0.048 indicates. Despite some of the sources in this sample appearing to be visually disturbed, none are identified as mergers by the $CA$ analysis. \citet{2000ApJ...529..886C} and \citet{2003ApJS..147....1C} show that while concentration is largely unaffected by increasing redshift, asymmetry is weakened to varying degrees, depending on the {\it HST} instrument used. For example, low redshift galaxies would typically have their asymmetries decreased by -0.10${\pm}$0.10 (for ACS) and -0.03${\pm}$0.10 (for WFPC2) if they were observed instead at $z$=2. These are not strong effects, but because the dark GRB host redshift distribution is shifted to slightly higher values with respect to \citet[][see figure \ref{fig:redshifts}]{2005ApJ...633...29C}, it is worth considering the dilution of asymmetry as an explanation for the lack of $CA$ mergers in this sample. 

\subsection{Host Galaxy -- GRB Offsets} 
As we showed in section \ref{sec:colmag}, the majority of sources in this sample are consistent with dusty galaxies lying at intermediate redshifts. Their host-normalised offset distributions are similar to those of optically-bright GRBs. This has implications for how the extinguishing dust is distributed within the galaxies. If the dust were uniformly spread throughout the internal volume of the galaxy, there would be a tendency for dark bursts to occur at low projected offsets, where the column density of dust would be greater. This assumes, however, that the underlying form of the star formation distribution is the same in both dusty and non-dusty GRB hosts, and that the nature of the dust is the same throughout the galaxy. There are also uncertainties on the true galaxy centroids arising from the blurring effect of dust, and the irregular nature of the galaxy morphology in some cases. Nevertheless, the fact that we do not see any bias to low offsets implies either that the extinction occurs in a foreground screen (difficult to arrange for every system) or that the dust has a clumpy component. This clumpiness is likely on galactic scales (hundreds to thousands of parsecs), not on the scale of the afterglow radiating region, as has been shown by studies of absorption lines in GRB afterglows \citep[e.g.][]{2008ApJ...685..344P,2012MNRAS.419.3039C} which put dense gas clouds at distances of a few hundred parsec from the burst (although we note these studies require optical afterglows and therefore do not use dark GRBs). Additionally, if the dust causing extinction were too close to the GRB and natal site of the progenitor, it would likely be destroyed \citep{2000ApJ...537..796W,2001ApJ...563..597F,2018MNRAS.479.1542Z,2018arXiv181011064H}. The inference that clumpy dust is present is further supported by the ellipticity distribution: assuming that at least some of the hosts are spiral in morphology, the lack of a favoured edge-on orientation suggests that the line-of-sight depth through the host is not the only factor in causing darkness in GRBs. 

There is a bias against the measurement of very small offsets when afterglow positional uncertainties are an appreciable fraction of the projected size of the galaxy \citep{2016ApJ...817..144B}. This issue is therefore more significant when {\it Chandra} X-ray positions are used, which although precise, are typically less so than optical, IR or radio localisations. One method of addressing this is to include more dark bursts with optical or NIR positions, as in section \ref{sec:burstoffsets}. While still statistically consistent with the comparison samples, we can see from figure \ref{fig:f160w_offsets} that the inclusion of extra optically/NIR detected GRBs shifts the dark offset distribution to smaller values, as expected. We caution that in our sample, only 3 dark bursts from 21 have an optical/NIR localisation - suggesting that such scenarios are rare. By artificially including more of these in our extended sample, we must acknowledge that selection biases might be introduced. However, the redshift and physical host sizes are not significantly changed by their inclusion (mean $z$ changes from 1.44 to 1.61, the mean R$_{50}$ from 2.67 to 2.40). It therefore appears as though such biases are not a large concern. Overall, the inclusion of extra optically/NIR detected dark GRBs does not change our interpretation of the results.

There is mixed support for a clumpy dust model in the literature. \citet{2018A&A...617A.141C} studied the relationship between line-of-sight extinction curves derived from optically-bright GRB afterglows and the global dust properties of the host. They find that for more than half of their sample, a significant amount of clumpy dust is required. 
We would certainly expect some level of irregularity in the dust distribution as both supernovae and mature stellar populations are dust-production sites and these will each enrich a limited volume. 
Since star-forming galaxies tend not to have uniform distributions of star formation, the starburst regions will randomly sample dusty and dust-sparse sites. Indeed, \citet{2011A&A...534A.108K} find examples of heavily extinguished bursts in otherwise blue, low mass galaxies, as might be expected given this model. \citet{2015MNRAS.451..167F} and \citet{2017A&A...601A..83H} also find evidence for local dust properties which differ from the galaxy wide average. Furthermore, as figure \ref{fig:zmags} demonstrates, there is considerable overlap in absolute magnitudes between dark and bright GRB hosts. Hosts of similar absolute magnitude are capable of producing both bright and dark bursts, which suggests that the host luminosity and/or mass does not correlate directly with the dustiness of sight-lines through the galaxy. There is also much uncertainty about the interstellar dust properties of GRB hosts \citep{2018MNRAS.479.1542Z,2018MNRAS.480..108Z}, and variations in $R_{V}$ between hosts and/or burst sites could help explain the overlap in dark and bright GRB host luminosities.

In argument against the clumpy dust scenario, the fact that hosts of the same absolute magnitude can produce both bright and dark GRBs could be explained by whether or not the burst occurs on the near or far side of the galaxy. Furthermore, previous studies have indicated that burst site properties such as A$_{V}$ and metallicity are more typically similar to the properties derived from integrating over the entire host \citep{2011A&A...534A.108K}. From our sample of 21, only 3 GRB hosts have metallicity determinations. These are GRBs\,051022 \citep[$Z{\sim}8.77$,][]{2015arXiv151100667G}, 090407 \citep[$Z{\sim}8.85$][]{2012ApJ...758...46K} and 100615A \citep[$Z{\sim}8.4$][]{2012ApJ...758...46K}. The first two of these have ${\sim}$ solar metallicity, which is particularly high for GRB hosts, and consistent with the presence of dust. This implies that, for the progenitors at least, the main difference between dark and bright GRBs may be their metallicity. The link between host and afterglow determined properties is by definition only measurable for bursts where the afterglow is detected, and it remains possible that the local A$_{V}$ might be greater than the host average for dark GRBs. Nontheless, \citet{2013ApJ...778..128P} and \citet{2016ApJ...817....8P} concluded that an approximately uniform dust component can help explain the dark burst preference for massive hosts.

The true picture is likely not to be as simple as either purely homogeneously-distributed gas and dust, or entirely clumpy. Some combination of these extremes, with clumps occurring embedded within more diffuse dust is most likely, in agreement with the findings of \cite{2018A&A...617A.141C}. Ultimately, a study of how ${\beta}_{\mathrm{OX}}$ varies with host normalised offset would be able to distinguish the various dust distribution scenarios discussed here.

\section{Conclusions}\label{sec:conc}
We present F606W and F160W imaging of a sample of 21 dark bursts, where the burst location is known in 20 cases to significantly sub-arcsecond (typically $\sim0\farcs1-0\farcs3$) precision through {\it CXO} X-ray afterglow observations. Twenty of the bursts are robustly detected in the F160W band, and twelve at F606W. One source is undetected in both bands.
Where sources are undetected, deep ${\it HST}$ imaging allows us to place stringent limits on  host magnitudes, and thus evaluate the plausibility of a high redshift interpretation for optically-faint afterglows. This analysis provides an upper limit of 22 per cent of dark GRBs arising from $z>5$, or ${\sim}$ 4.4 per cent of all GRBs, consistent with previous estimates. 

We also consider the morphology of the detected hosts.  A concentration and asymmetry analysis provides marginal evidence that dark GRB hosts are more concentrated than the hosts of optically-bright GRBs. Otherwise, the morphologies of these galaxies are consistent with the wider GRB host population. In agreement with previous studies, we have shown that dark gamma-ray bursts occur preferentially in galaxies which are larger and more luminous that those hosting optically bright bursts. Dark bursts trace their host light in a similar way to bright GRBs, with no evidence for a smaller offset bias. Combining ellipticities with the concentration and asymmetry parameters, we find that dark hosts do not show any evidence for a preferred edge-on orientation. This, and the offset distribution, may imply that a significant proportion of the extinguishing dust is clumpy on galactic scales.

\section*{Acknowledgements}
AAC is supported by STFC grant 1763016. AAC also thanks the William Edwards educational charity. AJL and PJW have been supported by STFC consolidated grant ST/P000495/1.

Based on observations made with the NASA/ESA Hubble Space Telescope, obtained
from the data archive at the Space Telescope Science Institute. STScI is operated by the Association of Universities for Research in Astronomy, Inc. under NASA contract NAS 5-26555. These observations are associated with programs GO 11343, 11840, 12378, 12764, 13117 and 13949 (Levan). Based on observations collected at the European Southern Observatory under ESO programme 095.B-0811(C). The scientific results reported in this article are based on observations made by the Chandra X-ray Observatory. This research has made use of software provided by the Chandra X-ray Center (CXC) in the application of the CIAO package.

The Pan-STARRS1 Surveys (PS1) and the PS1 public science archive have been made possible through contributions by the Institute for Astronomy, the University of Hawaii, the Pan-STARRS Project Office, the Max-Planck Society and its participating institutes, the Max Planck Institute for Astronomy, Heidelberg and the Max Planck Institute for Extraterrestrial Physics, Garching, The Johns Hopkins University, Durham University, the University of Edinburgh, the Queen's University Belfast, the Harvard-Smithsonian Center for Astrophysics, the Las Cumbres Observatory Global Telescope Network Incorporated, the National Central University of Taiwan, the Space Telescope Science Institute, the National Aeronautics and Space Administration under Grant No. NNX08AR22G issued through the Planetary Science Division of the NASA Science Mission Directorate, the National Science Foundation Grant No. AST-1238877, the University of Maryland, Eotvos Lorand University (ELTE), the Los Alamos National Laboratory, and the Gordon and Betty Moore Foundation.

IRAF is distributed  by  the  National  Optical  Astronomy  Observatory,  which  is  operated  by  the  Association  of Universities for Research in Astronomy (AURA) under cooperative agreement with the National Science Foundation. This research has made use of the SVO Filter Profile Service (\url{http://svo2.cab.inta-csic.es/theory/fps/}) supported from the Spanish MINECO through grant AyA2014-55216. We acknowledge the use of Ned Wright's online cosmology calculator \citep{2006PASP..118.1711W}.




\bibliographystyle{mnras}
\bibliography{darkgrbs} 


%
%


\bsp	
\label{lastpage}
\end{document}